\documentclass[journal=jacsat,manuscript=communication]{achemso}
\usepackage[version=3]{mhchem}

\usepackage{graphicx} 
\usepackage{subfig}
\usepackage{multirow}
\usepackage{longtable}
\usepackage{achemso} 
\usepackage{array,booktabs}
\usepackage{float}
\newcolumntype{C}[1]{>{\centering\arraybackslash}p{#1}}
\usepackage{caption}
\usepackage{color}
\usepackage[]{siunitx}
\usepackage{verbatim} 
\setkeys{acs}{usetitle = true}
\usepackage{color, colortbl}
\definecolor{Gray}{gray}{0.9}

\author{Tarak Karmakar$^{\dagger,\ddagger}$, Pablo M. Piaggi$^{\dagger,\ddagger}$ and Michele Parrinello$^{\dagger,\ddagger}$}
\email{michele.parrinello@phys.chem.ethz.ch}
\vspace{20mm}
\affiliation{$^\dagger$Department of Chemistry and Applied Biosciences, ETH
Z\"{u}rich, c/o USI Campus, Via Giuseppe Buffi 13, CH-6900, Lugano, Ticino,
Switzerland\\
$^\ddagger$Facolt\`{a} di Informatica, Istituto di Scienze Computationali,
Universit\`{a} della Svizzera Italiana (USI), Via Giuseppe Buffi 13, CH-6900,
Lugano, Ticino, Switzerland.\\}

\title[\texttt{achemso}]
{Simulations of Crystal Nucleation from Solution at Constant Chemical Potential}

\begin{document}

\onecolumn
\maketitle
\begin{abstract}
A widely spread method of crystal preparation is to precipitate it from a supersaturated solution. In such a process, control of solution concentration is of paramount importance. Nucleation process, polymorph selection, and crystal habits depend crucially on this thermodynamic parameter. When performing simulations in the canonical ensemble as the crystalline phase is deposited the solution is depleted of solutes. This unavoidable modification of the thermodynamic conditions leads to significant artifact. Here we adopt the idea of the constant chemical potential molecular dynamics approach of Perego {\em et al. [J. Chem. Phys. 2015, 142, 144113]} to the study of nucleation. Our method allows determining the crystal nucleus size and nucleation rates at constant supersaturation. As an example we study the homogeneous nucleation of sodium chloride from its supersaturated aqueous solution.

\end{abstract}

\twocolumn
\newpage
\noindent
{\bf 1. INTRODUCTION}\\

Crystal nucleation and growth from solution have great impact in chemical, material, biological, and environmental sciences. In solution, crystallization
occurs by the aggregation of molecules or in general particles to
form nuclei followed by their growth into macroscopic crystals. The early stage of this process contains valuable information on the microscopic pathways
that lead to the formation of the crystal. However, unveiling such details of this deceivingly simple process is a challenging task. Experiments have great difficulty in resolving the length and time scale of this process, and simulations have been proven to be of great help in this respect.~\cite{ohtaki1991nucleation,anwar1998computer,shore2000simulations,sarupria2012homogeneous,salvalaglio2015molecular,sosso2016crystal,fitzner2017pre,sosso2018unravelling} 

In crystallization from solution, supersaturation plays a crucial 
role in determining nucleation mechanisms~\cite{vekilov2010two,myerson2013nucleation,de2013crystal,tan2016understanding,lee2016multiple} and modulating polymorph selection~\cite{sudha2013supersat,liang2016supersat,liu2017supersat}. Thus it is 
important to study nucleation at constant solution supersaturation condition. However, modeling such a process on the computer is a non-trivial exercise. Computer simulations of nucleation from solution suffer from several limitations. One such limitation is the finite-size effect that arises while simulating nucleation in a small system with a fixed number of particles (typically a few thousands). During nucleation and growth, solute molecules are continuously drawn from the solution, and contrary to experiments, the finite-sized model system fails to retain a steady solution concentration in front of the growing nucleus. This solution depletion drastically affects further growth. Many remedies have been proposed to address this issue. Among them, the simplest one is to add an analytical correction term to the free energy profile.~\cite{salvalaglio2012uncovering,agarwal2014nucleation}
The other option is to simulate a considerably large system in which the 
finite-size effect is negligible.~\citep{wedekind2006finite} However, simulating a large system is often cumbersome. Alternatively, one could simulate
an open system that can exchange particles with a fictitious external reservoir.~\cite{horsch2009grand} However, in many cases especially dense fluids, these methods encounter limitations due to low acceptance probabilities of particle insertion and deletion steps. Liu {\em et al.} proposed a string method in the osmotic ensemble to carry out constant supersaturation simulations.~\cite{liu2018modelling} 

A more direct approach, called constant chemical potential molecular dynamics (C$\mu$MD) has been recently introduced by Perego {\em et al.} in which the solution concentration in contact with a growing crystal slab is maintained at constant supersaturation leading to steady crystal growth.~\cite{perego2015molecular} The C$\mu$MD method~\cite{perego2015molecular} and its cannibalistic variant~\cite{karmakar2018cannibalistic} were used to investigate crystal growth and dissolution of organic molecules and active pharmaceuticals from solution at different supersaturations~\cite{bjelobrk2019naphthalene} and various solvents~\cite{han2018solvent}. 

In the C$\mu$MD approach, a slab geometry was adopted which was suitable for studying  growth.~\cite{perego2015molecular} However to study a nucleation process an isotropic model is more appropriate. Here, we present such a variant designed to carry out simulations of crystal nucleation from solution at constant supersaturation. In our method, a nucleus is grown inside a sphere, and the solution concentration inside a shell surrounding the sphere is maintained at a target concentration. We have applied our method to study the homogeneous nucleation of sodium chloride (NaCl) from aqueous solution at constant supersaturation. The reasons for the choice of this system are two-fold; first, it is a challenging multi-component system, and secondly, the availability of a substantial amount of simulation results~\cite{zahn2004atomistic,zimmermann2015nucleation,lanaro2016birth,zimmermann2018nacl,jiang2018communication,jiang2018forward,jiang2019nucleation,patel2019simulations} makes it an excellent system to test new methods.\\

\noindent
{\bf 2. COMPUTATIONAL METHODS}\\

\begin{figure}
 \includegraphics[scale=2.5]{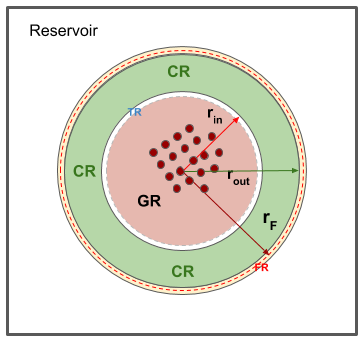}
 \caption{A two dimensional schematic description of the model system. The
	GR is shown in transparent red sphere. The white shell surrounding 
	GR is the TR. The green shell is the CR. The force region (width {\em
	w$_{F}$}) is depicted by the yellow shell. The red dotted circle
	indicates the force center ({\em $r_{F}$}). A representative nucleus is
	shown in the GR by a cluster of red circles.}
\label{fig:schem} 
\end{figure}

\noindent
{\bf{\em Constant Chemical Potential Simulation}}\\
In our C$\mu$MD scheme, a sphere and a set of concentric shells are defined as shown schematically in Fig. {\ref{fig:schem}}. A spherical region of a fixed radius is selected from the simulation box center. This region is called the growth region (GR). The nucleus is grown in this region. The GR is chosen large enough to accommodate a nucleus larger than the critical nucleus size. A thin transition region (TR) shell is defined outside of the GR. A shell outside of the TR is defined by choosing inner and outer boundaries, $r_{in}$ and $r_{out}$, respectively (Fig. {\ref{fig:schem}}). This shell is called the control region (CR). The solution concentration in this shell is maintained at a target concentration by an applied external force to be discussed later. The force-region (FR, yellow shell in Fig. 1) can be thought of as a membrane that allows solutes to enter/leave the CR depending on the concentration drop/increase in the CR. The region outside the FR serves as a molecular reservoir (buffer region) that, whenever needed, supplies solute atoms to the CR and thereby to the GR. The reservoir and the nested shells are periodically replicated, and periodic
boundary conditions are imposed. 

The solute concentration ($c$) in the CR is calculated as,
\begin{equation}
c = \frac{1}{V^{CR}} \sum^{N}_{j=1} f(r_j),\\
\label{eq:nCR}
\end{equation}
where $V^{CR}$ is the CR volume and $r_j$ the distance of a j-th particle from the box center. The $f(r_j)$ is a continuous and differentiable switching function that
counts atoms belonging the CR. In our case, this function is a defined as a product of two Fermi switching functions, $f_{in}$ (inward) and $f_{out}$ (outward), 
\begin{equation}
\begin{split}
f(r_j) &= f_{in}(r_j). f_{out}(r_j)\\
       &= \frac{1}{1+e^{-(r_j-r_{in})/\alpha}}.\frac{1}{1+e^{(r_j-r_{out})/\alpha}}
\end{split}
\label{eq:fermi}
\end{equation}
where, $r_{in}$ and $r_{out}$ are the inner and outer CR boundaries (Fig. \ref{fig:schem}), respectively, and $\alpha$ is a parameter that controls the switching functions steepness. The function $f(r_j)$ has a value of 1 when a solute is inside the CR and continuously approaches zero when outside. From now onward, we drop the atomic index for simplicity.

The force that restraints the instantaneous solution concentration ($c$ in Eq. \ref{eq:nCR}) at a target value ({\it $c_0$}) has the following expression, 
\begin{equation}
F(r) = \kappa(c-c_{0})G(r)
\label{eq:F}
\end{equation}
where $\kappa$ is the force constant. The $G(r)$ in Eq. (3) is a
bell-shaped function that localizes the {\it $F(r)$} at $r_F$. 
This function is defined as, \begin{equation}
	G = \frac{1}{2\sigma}\left [ \frac{1}{1+cosh(\frac{r-r_F}{\sigma})} \right ]
\end{equation}
where $\sigma$ is a broadening parameter. Details of the C$\mu$MD protocol 
 and parameters used in our simulations are provided in Section 1 the SI.\\

\noindent
{\bf {\em Collective variables (CVs)}}\\
Crystal nucleation is a rare event that occurs on the time scale difficult to
reach by regular atomistic simulations. Thanks to enhanced sampling
simulation methods, we can circumvent the time scale limitation and study
many long time-scale chemical and biophysical processes in short simulations at an affordable computational cost. One such enhanced sampling technique that has been used in many fields and demonstrated to be rigorous is metadynamics,~\cite{laio2002escaping,laio2008metadynamics,barducci2011metadynamics} especially in its well-tempered (WTMetaD) version.~\cite{barducci2008well,dama2014well} In this method, a history-dependent external bias is constructed as a function of a set of CVs that are a function of the atomic coordinates. Application of such a bias potential allows the system to transform from one state to another, which in the context of nucleation, are the solution without and with a crystal nucleus. In the following section, we provide details of the collective variables that we have used in the WTMetaD simulations. We introduce two CVs, one related to the local crystalline order and the other to the ionic solvation.\\

\noindent
{\em Local Order ($CV_1$)}\\
The first CV is based on the local ordering of $n$ neighbors ($i$) of a central atom in a crystal environment, $\chi$.  The local density $\rho_{\chi}(\bf r)$ of the central atom is written as a sum of Gaussian functions,
\begin{equation}
\begin{split}
\rho_\chi(\bf r) &= \sum_{i\in \chi} e^{-|{\bf r_i} - {\bf r}|^2/2\sigma^2}
\end{split}
\label{eq:rho}
\end{equation}
where ${\bf r_i}$'s are the coordinates of the neighbors relative to the central atom. $\sigma^2$ is the variance of the Gaussian functions. We now define 
a reference crystal environment ($\chi_0$) and choose $n$ nearest neighbors
positions {$\{\bf r_i^0\}$} of the central atom in the crystal lattice. The difference between the two environments $\chi$ and $\chi_0$ is then calculated by a kernel 
function as done in ref.~\citenum{bartok2013representing, de2016comparing,piaggi2019phase},
\begin{equation}
\begin{split}
k_{\chi_0}(\chi) &= \int dr \rho_\chi(\bf r) \rho_{\chi_0}(\bf r)
\end{split}
\label{eq:int}
\end{equation}
where $\rho_{\chi_0}(\bf r)$ is the local density of the atom in the reference crystal environment ($\chi_0$). In ref~\citenum{de2016comparing}, the reference
environment is not fixed in space but rotated so as to obtain a rotationally 
invariant CV. Here we break the symmetry. In doing so, the CV acquires a simple analytical expression.~\cite{piaggi2019phase}
\begin{equation}
\begin{split}
k_{\chi_0}(\chi) &= \sum_{i\in \chi} \sum_{j\in \chi_0} \pi^{3/2} \sigma^3 e^{-|{\bf r_i} - {\bf r_j}|^2/4\sigma^2}
\end{split}
\label{eq:kernel}
\end{equation}
One shortcoming of this CV is that it can change its value by an overall
rotation of the crystal. Since this is artificial, we add a restraint in order
to avoid this unwanted effect (details can be found in SI).

The kernel function in Eq. (\ref{eq:kernel}) is then normalized such that similarities between the identical environments e.g., $\bar{k}_{\chi}(\chi)$ and $\bar{k}_{\chi_0}(\chi_0)$ are equal to one. 
\begin{equation}
\begin{split}
\bar{k}{_{\chi_0}(\chi)} &= \frac{k_{\chi}(\chi_0)}{k_{\chi_0}(\chi_0))}\\
           &= \frac{1}{n}\sum_{i\in \chi} \sum_{j\in \chi_0} e^{-|{\bf r_i} - {\bf r_j}|^2/4\sigma^2}
\end{split}
\label{eq:nkernel}
\end{equation}

Now let us consider a system containing $N$ solute particles. For each 
solute ($i$ = 1,..,$N$) we calculate the kernel function $\bar{k}_{\chi_0}(\chi_i)$ using Eq. (\ref{eq:nkernel}). The solutes having $\bar{k}_{\chi_0}(\chi_i)$ $>$ ${k_0}$, where ${k_0}$=0.5, are counted using a continuous
and differentiable switching function ($s^O_{i}$) as follows,
\begin{equation}
\begin{split}
s^O_{i} &= \frac{1-(\bar{k}_{\chi_0}(\chi_i)/k_0)^p}{1-(\bar{k}_{\chi_0}(\chi_i)/k_0)^q}
\end{split}
\label{eq:cv1}
\end{equation}
The variable $s^O_{i}$ has values in the range from 1 to 0; 
for crystalline atoms in a perfect environment, $s^O_{i}$ $\approx$ 1 while those in solution, $s^O_{i}$ $\approx$ 0. The $s^O_{i}$ can be referred to 
as an atomic crystalline CV. Here the superscript {\it O} refers to the local order. The parameters p and q control the steepness and the range of the switching function. 

The compound NaCl has a rock-salt crystal structure and the ions Na$^+$ and Cl$^-$
form two interpenetrating FCC lattices. Each ion is surrounded by six of the oppositely charged ion arranged in an octahedral symmetry. We use one of these local environments as a reference structure. Here we have used as reference Na$^+$ and its Cl$^-$ neighbors. However we could as well have used Cl$^-$ as the central atom. A lattice parameter of 0.282 nm with $\sigma$ = 0.08 nm was used for the kernel defined in Eq. (\ref{eq:nkernel}). 

Furthermore, since we were interested in biasing only the Na$^+$ ions inside the GR (Fig. \ref{fig:schem}), we used another switching function, $\omega (r_i)$ acting on the distance ($r_i$) between the $s^O_{i}$ (Eq. \ref{eq:cv1})  CVs position and the reference simulation box center, and to measure whether the $s^O_{i}$s are inside the sphere or not. Finally, we define our first CV ($s^O$) as the sum over the crystalline Na$^+$ ions that are present inside the sphere, 
\begin{equation}
s^O = \sum_{i} s^O_{i} \omega(r_i)
\label{eq:scv1}
\end{equation}
The switching function ($\omega(r_i)$) decays smoothly from 1 at a radius 1 nm to 0 beyond 1.5 nm from the box center.\\

\noindent
{\em Ion hydration ($CV_2$)}\\
Solvent plays a pivotal role in the nucleation of ionic salts. During nucleation, increase in solute density is accompanied by ion dehydration; the latter is often found to be rate determining step.~\cite{zahn2004atomistic,piana2006assisted,kowacz2007effect,raiteri2010water,joswiak2018ion} In aqueous solution, Na$^+$ and Cl$^-$ ions are solvated with an average coordination number 6 and 8, respectively. In order to accelerate ions dehydration, we introduced another CV ($s^H_{i}$) which is based on water coordination number of each Na$^+$ ion ($i$ = 1,..,$N$) within a given cut-off radius ($r_0$), 
\begin{equation}
\begin{split}
s^H_{i} &= \sum_{j}^{N_w} \frac{1-(r_{ij}/r_0)^p}{1-(r_{ij}/r_0)^q}
\end{split}
\label{eq:cv2}
\end{equation}
where, $s^H_{i}$ is the water coordination number of $i$-th Na$^+$ ion, and $r_{ij}$ and $r_0$ are the Na$^+$-O($H_2O$) distances and the distance cut-off, respectively. We have chosen $r_0$=0.4 nm in our case. $N_w$ is the total number of water molecules in the system. The parameters p and q control the steepness of the switching function.

Finally, the second CV ($s^H$) is a weighted-average water-coordination number of Na$^+$ ions inside the sphere and defined as,
\begin{equation}
s^H = \frac{\sum_{i} s^H_{i} \omega(r_i)}{\sum_i \omega(r_i)}
\label{eq:scv2}
\end{equation}
In this case also, a cubic switching function ($\omega(r_i)$) is used. We drive explicitly Na$^+$ solvation. Adding the Cl$^-$ solvation led to an increase in computational cost without much new insight.

It must be noted that by changing the geometry in Fig. 1 we cannot form infinitely repeated periodic crystals. For this reason we limit the size of the crystal that can be formed by imposing a restraint. We also found useful to put a limit to the number of waters that can solvate ions in order to avoid sampling configurations that are not relevant for the process under study.\\

\noindent
{\bf {\em System setup and simulation method}}\\
We considered a simulation box of dimension $\sim$ 6x6x6 nm$^3$. The box size
was sufficient to accommodate the critical nucleus and a crystal in the growth region. The system contains 1000 ion-pairs (1000 Na$^+$ and 1000 Cl$^-$) and 6000 water molecules. The solution concentration is $\sim$ 10 $m$ (mole of solutes per kg of solvent). A similar concentration has been considered in previous studies to simulate NaCl nucleation from aqueous solution.~\cite{zimmermann2015nucleation,jiang2018forward} The Juang-Cheatam force-field~\cite{joung2009molecular} was used to model the ions while water molecules were described with the SPC/E potential.~\cite{berendsen1987missing} A time-step of 2 fs was employed.  A cut-off of 0.9 nm was used for both van der Waals and short-range Coulomb interactions. The long-range electrostatic interactions were treated by the Particle Mesh Ewald method.~\cite{essmann1995smooth} 
The enhanced sampling runs were started after equilibration run of 5 ns. During this initial phase the pressure was set at 1 bar using the Parrinello-Rahman barostat~\cite{parrinello1981polymorphic}, while the temperature was kept at 350 K using the stochastic velocity rescaling thermostat.~\cite{bussi2007canonical} In the enhanced sampling runs that followed, we continued controlling the temperature with the same thermostat while the volume was kept constant thus switching to an (N,V,T) ensemble.

We have used the well-tempered metadynamics (WTMetaD) method to carry out nucleation simulations. In the WTMetaD simulations, an initial hill height of 30 kJ/mol and widths of 0.5 and 0.1 for $s^O$ and $s^H$, respectively were used.  Hills were deposited every 500 steps. A bias factor of 100 and 50 were used. 

All simulations were carried out using the GROMACS-2018.3 software patched with the PLUMED2 code. The C$\mu$MD method and CV related codes are included in a private version of the PLUMED2 plugin. A representative PLUMED input file containing the CVs' information and the metadynamics protocol is provided in Section 2 (Fig. S1-4) of the SI. The visual Molecular Dynamics (VMD) software~\cite{humphrey1996vmd} was used to visualize the trajectories and produce some of the figures.\\

\noindent
{\bf 3. RESULTS AND DISCUSSION}\\

After equilibration, we carried out three independent simulations (A, B, and C). In A, a standard NVT ensemble was used, and the initial value of the solution concentration was $c$ =  4.2 nm$^{-3}$. In B, the same concentration was imposed during the entire simulation run using the C$\mu$MD method. In C, the concentration was kept at the higher value of $c$ = 5.0 nm$^{-3}$ using again the C$\mu$MD approach. Contrasting A and B allows us studying the effect of keeping $c$ constant, while from the difference between B and C the effect of increasing $c$ can be investigated.

Let us start by discussing the results obtained from simulation A. When studying A even if not necessary, we use the CVs, $s^O$ and $s^H$ whose action is localized in a sphere of size 1.5 nm. This will allow a fairer comparison with a parallel C$\mu$MD calculation. In order to understand the effect of  nucleus growth on the solution concentration, we calculated its value in a shell 1.7 nm away from the box center and with a  thickness of 0.8 nm. This region is the same as the one that is used in C$\mu$MD simulations to monitor the solution concentration. During the metadynamics runs the concentration decreases as the size of the crystalline nucleus increases (see Fig. S6 of the SI). In contrast, in B and C, the instantaneous solution concentration fluctuates around the desired values ($c_0$) (Fig. S7). This accurate control of solution concentration demonstrates the effectiveness of our method.  
\begin{figure*}
  \begin{center}
    \includegraphics[width=0.9\textwidth]{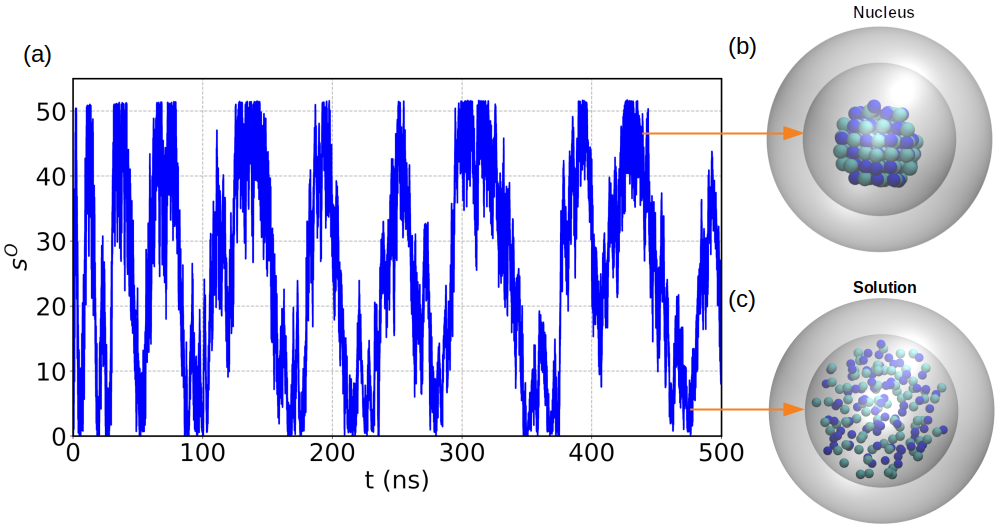}
  \end{center}
  \caption{(a) The $s^O$ CV plotted as a function of metadynamics
  simulation time (C simulation). The $s^O$ and $s^H$ profiles obtained from A and B simulations are provided in Fig. S5 of the SI. Representative frames extracted from two states, (b) a nucleus and (c) homogeneous solution in the GR. In (b), only the ions which are part of the nucleus are depicted in large spheres, and in both figures, other atoms in the whole system are not shown for clarity.}
  \label{fig:n5res}
\end{figure*}

\begin{figure*}
  \begin{center}
    \includegraphics[width=1.0\textwidth]{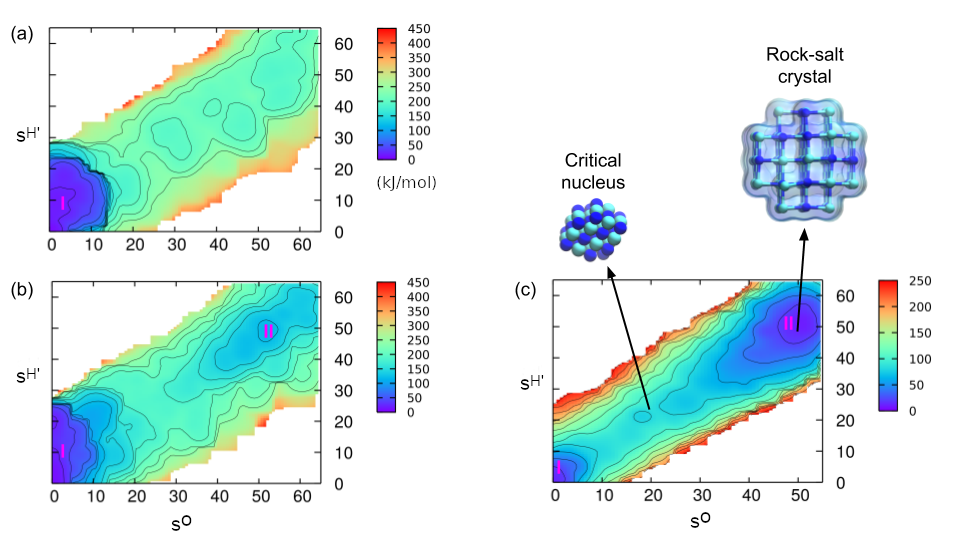}
  \end{center}
  \caption{Free energy surfaces as a function of $s^O$ and $s^{H'}$ obtained from the reweighting calculations on (a) A, (b) B, and (c) C simulation trajectories.     The solution and the crystal basins on the FESs are marked by the symbols I and II, respectively. A representative critical nucleus and a rock-salt crystal collected from the crystal basin (from the C simulation trajectory) are presented.}
  \label{fig:fes}
\end{figure*}

Now that the solution concentration is well-controlled in all cases, we shift our attention toward the nucleation events. Metadynamics is able to induce multiple transitions to and from a NaCl microcrystal (Fig. \ref{fig:n5res}(a) and Fig. S5). This enables us to collect enough statistics and calculate the free energy surfaces (FES). We followed a reweighting procedure discussed in ref.~\citenum{tiwary2014time}
to calculate the FESs as a function of $s^O$ and $s^{H'}$ CVs. Here, $s^{H'}$
is defined as the number of Na$^+$ ions having less than 3 water molecules within radius 0.4 nm (see section 6 of the SI for details of the reweighting protocol).  While the $s^O$ reveals local ordering of the ions, $s^{H'}$ unveils the solvation effect. In Fig. \ref{fig:fes}(a) and \ref{fig:fes}(b), we can clearly see the effect of controlling $c$. In \ref{fig:fes}(a), the absence of a crystal minimum is to be noted. In contrast in \ref{fig:fes}(b), the crystal state appears as a local minimum. Increasing the solution concentration stabilizes the crystal phase (Fig. \ref{fig:fes}(c)). Furthermore, an almost straight diagonal path for the phase transition indicates that there is a single path for the transformation from the solution to the crystal state demonstrating that during nucleation ions crystallization and desolvation are correlated.  

\begin{figure}[!ht]
  \begin{center}
    \includegraphics[width=0.45\textwidth]{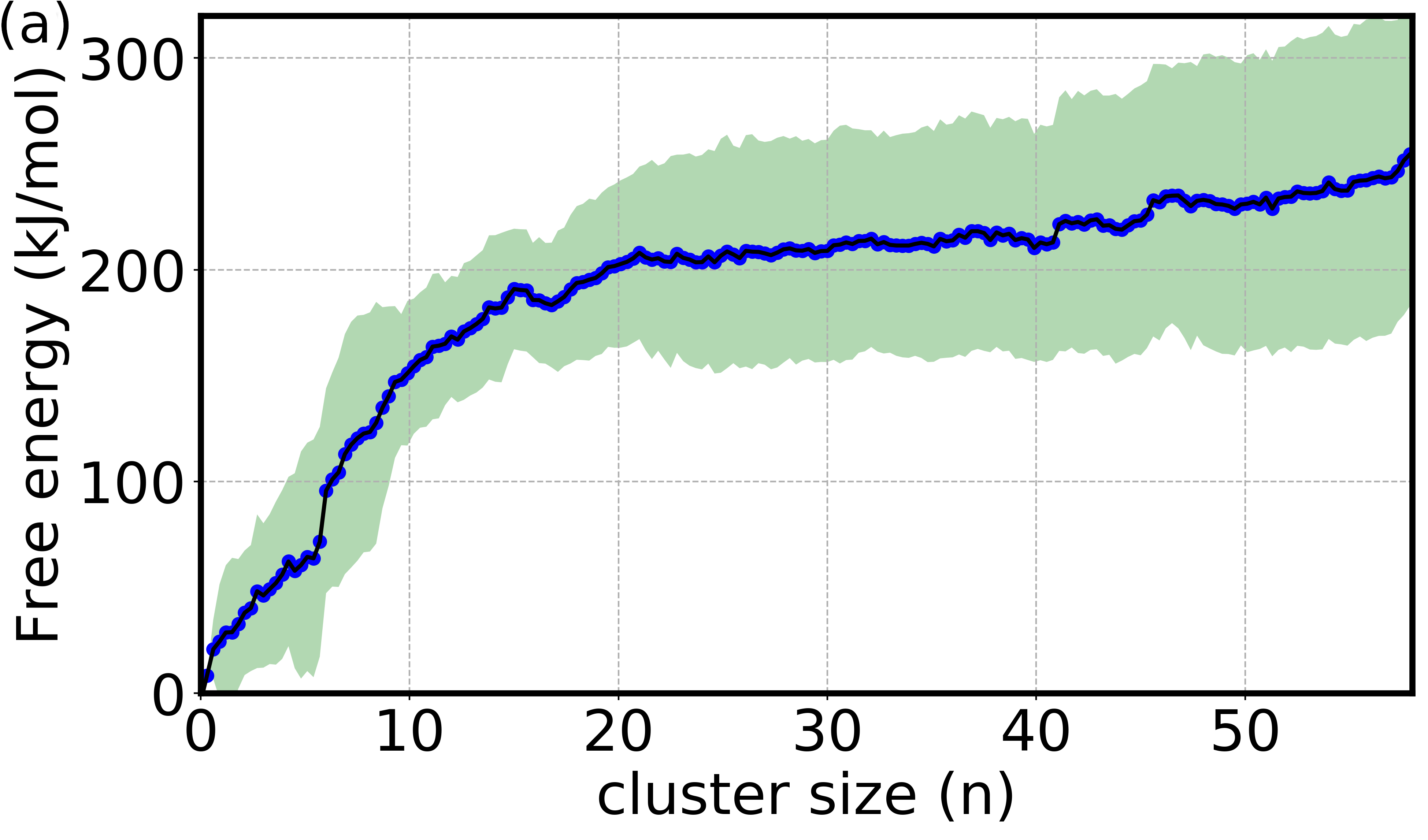}
    \includegraphics[width=0.45\textwidth]{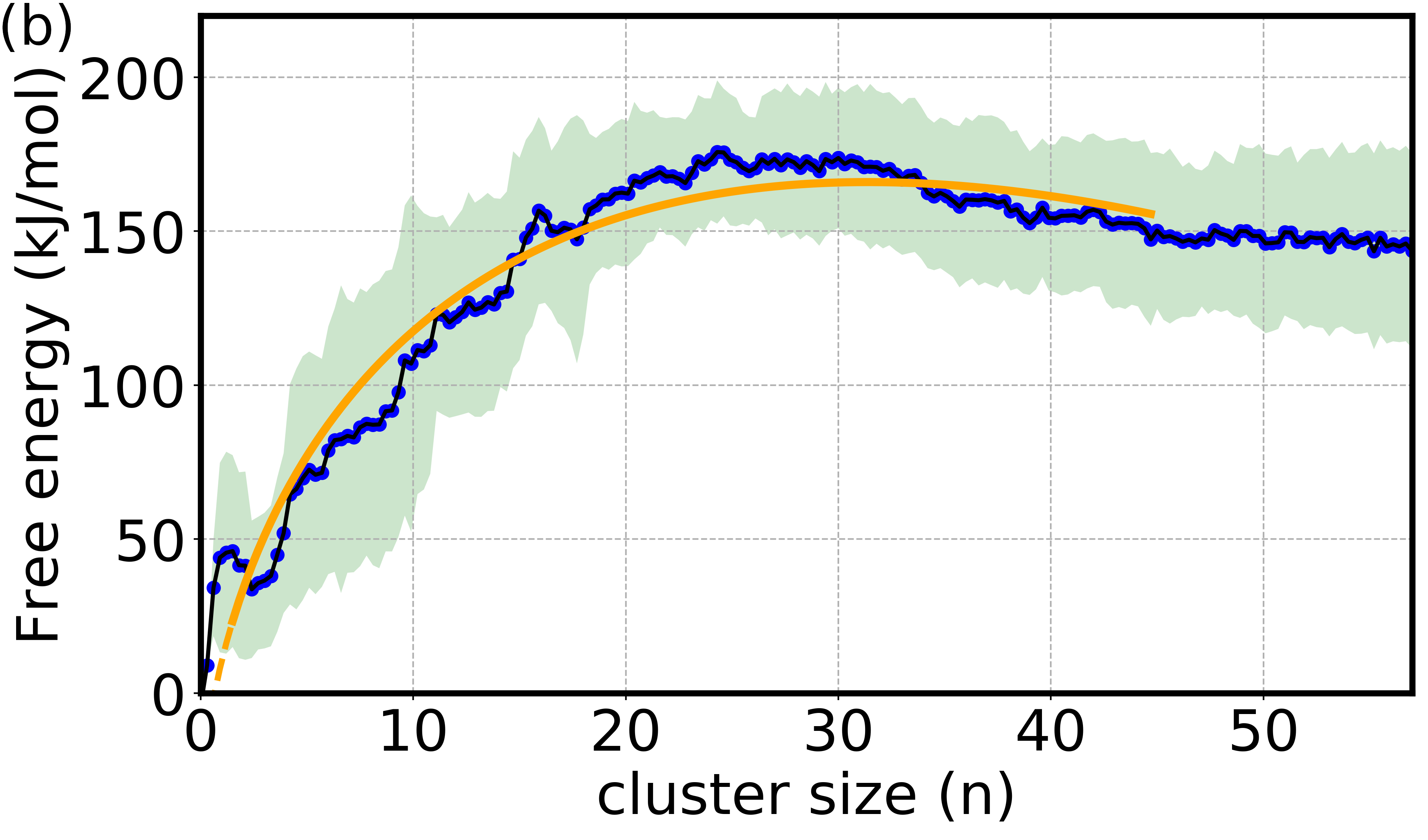}
    \includegraphics[width=0.45\textwidth]{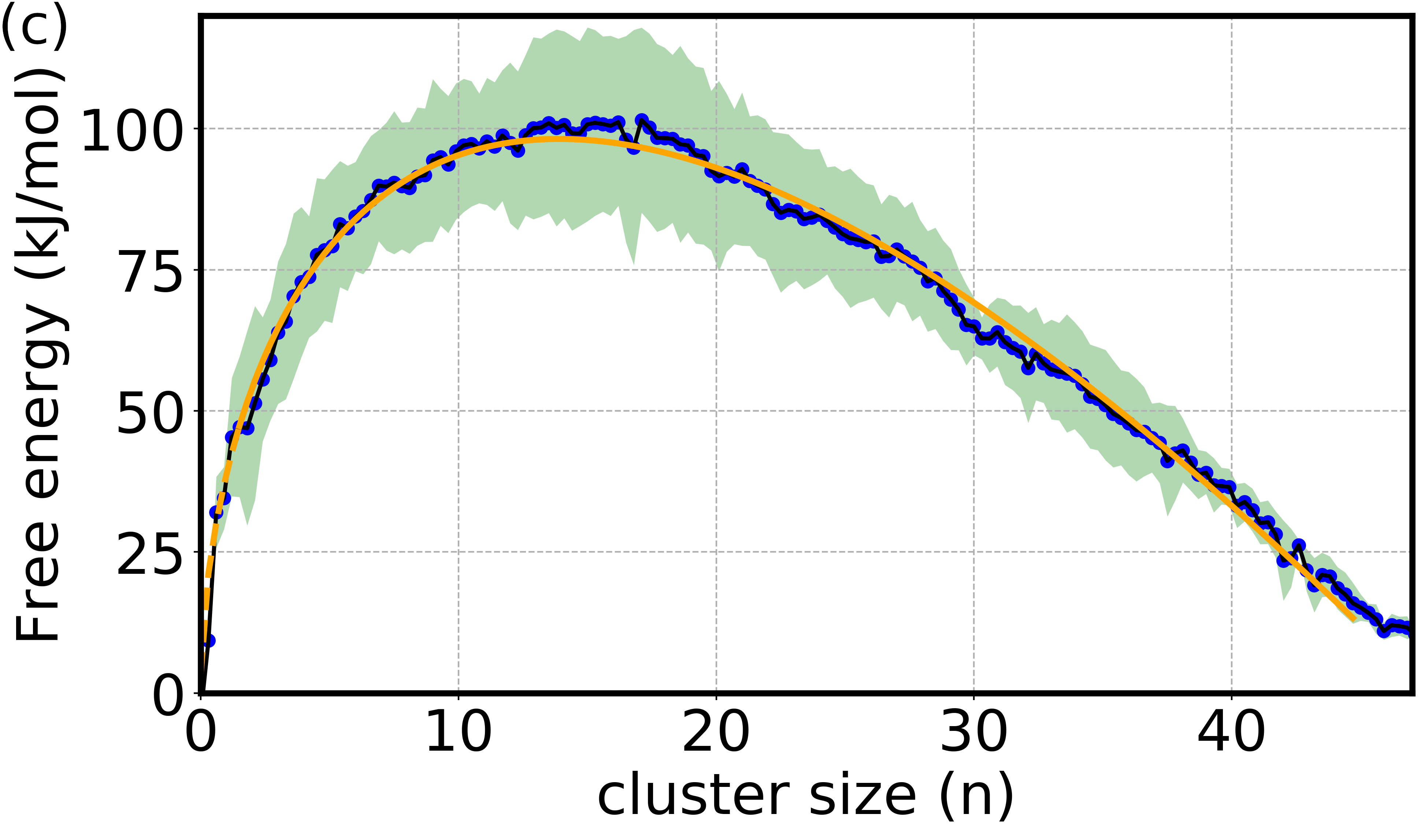}
  \end{center}
  \caption{Reweighted one-dimensional free energy profiles as a function of cluster size ($n$) obtained from (a) A, (b) B, and (c) C simulations, respectively. To calculate the errors in the free energy profiles, blocks of 100 ns width (50-150, 100-200,..., 400-500) were selected from the 450 ns simulation trajectory (first 50 ns of the trajectory was discarded), and the free energy in each block was calculated. The average free energy profile is shown in blue spheres, and the standard deviations in green filled-region.  The CNT-fitting curve is depicted in  orange line.}
  \label{fig:fep}
\end{figure}

Additionally, we calculated one dimensional free energy profiles from 
the cluster size (n) distributions as described in ref.~\citenum{piaggi2017variational,niu2018molecular} and~\citenum{niu2019temperature}. The average free energy profiles and the errors were calculated using the block averaging analysis. The free energy profile obtained from simulation A (Fig. \ref{fig:fep}(a)) increases monotonously as the nucleus size increases. This could be due to the varying thermodynamical driving force induced by the solution depletion. In the B simulation, the free energy shows a barrier of 176$\pm$28 kJ/mol, and after the critical nucleus, it decreases as the nucleus size increases. The lowering of the nucleation barrier in B simulation could be attributed to the controlled supersaturation condition. The critical nucleus ($N_c$) has been found to be comprised of approximately 27 ion-pairs. Furthermore, in the C simulation, with its increase in solution concentration, the free energy barrier decreases to 101$\pm$10 kJ/mol. In this case, a smaller critical nucleus of size $\sim$ 16 ion-pairs is obtained. The shape of these free energy curves (Fig \ref{fig:fep}(b) and (c)) spurred us to check if they fit to the classical nucleation theory (CNT) curve, and in fact, not surprisingly, a good fitting is obtained. The fitting of the free energy curve to the CNT-equation provided the chemical potential difference $\Delta\mu$ = 13.0 kJ/mol for B and 13.8 kJ/mol for C simulation.

In the next step, following a protocol described in refs.~\citenum{becker1935kinetische,auer2001prediction,lundrigan2009test,sanz2013homogeneous,espinosa2016time,espinosa2016seeding,espinosa2014homogeneous,soria2018simulation} we calculate the nucleation rate ({\em J}) 
\begin{equation}
    J = \rho Z f^+ exp (-\Delta F/k_BT)
    \label{eq:rateJ}
\end{equation}
where $\rho$ is the number density, $Z$ the Zeldovich factor, $f^+$ the attachment frequency, $\Delta F$ the nucleation barrier, $k_B$ the Boltzmann constant, and T the temperature. The $Z = \sqrt{|\Delta \mu|/6 \pi k_B T N_c}$ is calculated using $\Delta\mu$ and $N_c$ values. The attachment frequency ($f^+$) was obtained from the average slope of the $<(N-N_c)^2>$ vs. time curves obtained from a few short unbiased simulations initiated from configurations containing a critical nucleus (Fig. S11). All values related to the rate calculation are provided in Table S1 of the SI. Finally, we inserted all these values to the rate equation (Eq. \ref{eq:rateJ}) and obtain  nucleation rates of 2.5$\times$10$^{3}$ cm$^{-3}$s$^{-1}$  and 6.2$\times$10$^{14}$ cm$^{-3}$s$^{-1}$ at solution concentration $c$ = 4.2 nm$^{-3}$ and $c$ = 5.0 nm$^{-3}$, respectively, and at temperature 350 K. The unavailability of NaCl nucleation rate at high temperatures does not allow a direct comparison of our results with experiments. However, the obtained nucleation rates are close to the values (10$^1$ - 10$^{25}$ cm$^{-3}$s$^{-1}$) at ambient conditions.~\cite{na1994cluster,gao2007efflorescence,jiang2018forward,zimmermann2018nacl}

Although not the primary focus of our work, here we provide a summary of 
observations related to the nucleation mechanism gathered from the simulation trajectories. A visual inspection of the simulation trajectories reveals single-step nucleation mechanism. Moreover, the CNT-like free energy profile obtained from the C$\mu$MD simulations depicted in Fig. \ref{fig:fep}(b) and (c) is an indicator of a likely one-step nucleation mechanism. Our observations are in line with the previous reports.~\cite{jiang2018forward,jiang2019nucleation} Furthermore, in our simulations, we do not observe the Wurtzite structure reported in a few studies.~\cite{giberti2013transient,zimmermann2015nucleation} This could be due to the choice of our CV which only describes the rock-salt structure. A movie demonstrating a representative nucleation event extracted from the C simulation is provided as a web-enhanced SI.\\

\noindent
{\bf 4. CONCLUSION}\\

In summary, we have developed a C$\mu$MD simulation method to carry out simulation of nucleation from solution at constant chemical potential. 
The presented method has been demonstrated to be effective in controlling
the solution concentration near the growing nucleus. Active control of the solution supersaturation surrounding the growing embryo leads to steady nucleation. Although in this particular case, the solution depletion is not severe ($\sim$7-10 \%), it could be more significant when a large crystal is grown from solution in a closed finite-size system.  

An effective chemical potential control together with metadynamics simulations helped us in calculating NaCl nucleation free energy surfaces at a given solution concentration. The local order-based collective variable ($s^O$) employed in our metadynamics simulations has been found to be delicate in growing a target crystal structure.  A single pathway for the transformation from the solution to the crystal has been realized. The CNT-like free energy curve further confirms that NaCl follows a single-step nucleation in supersaturated aqueous solution.  

The C$\mu$MD method introduced here opens up many possible applications. An immediate
application would be to carry out seeded nucleation at constant solution concentration. Alternatively one can interface our method with several enhanced sampling methods~\cite{filion2010crystal,vanden2010transition} focusing nucleation e.g., forward-flux sampling~\cite{allen2009forward}, persistent-embryo approach~\cite{sun2018overcoming}, and string method~\cite{weinan2005finite,maragliano2006string,liu2018modelling} to carry out controlled nucleation at constant solution concentration. Furthermore, multi-resolution scheme such as adaptive resolution~\cite{praprotnik2008multiscale,praprotnik2005adaptive,praprotnik2006adaptive,fritsch2012adaptive,delle2017molecular,praprotnik2018adaptive,krekeler2018adaptive,ciccotti2019physics} can also be combined with the presented C$\mu$MD technique to carry out concentration-controlled nucleation simulations of a system with a relatively large reservoir containing low-resolution coarse-grain solute and solvents.  

The method is no way limited to the study of nucleation but could be 
useful in simulating controlled self-assembly of small organic and biomolecules such as peptides. We believe the realistic nature of our method in mimicking bulk-like environment to the growing nucleus or cluster will help in obtaining 
true kinetics.\\

\noindent
{\bf 5. ASSOCIATED CONTENT}\\

\noindent
{\bf Supporting Information}\\
The supporting information contains C$\mu$MD protocol, sample PLUMED input files,
details of the restraints used in metadynamics simulations, time series of $s^O$ and $s^H$ CVs for A, B, and C simulations, a correlation plot of solution concentration and $s^O$ CV from A simulation, solution concentration profiles obtained from simulations B and C, time evolution of mean square change in the cluster size, a table containing values related to rate calculations.\\

A movie demonstrating a nucleation event can be found here ( )\\

The codes used in this work are included in the private development version
of PLUMED2 plugin and will be openly available in the future. Until the official
release, the codes will be available upon request to the corresponding
authors.\\

\noindent
{\bf 6. AUTHOR INFORMATION}\\

\noindent
{\bf Corresponding Author*}\\ 
E-mail: michele.parrinello@phys.chem.ethz.ch\\

\noindent
{\bf ORCID:}\\
Tarak Karmakar: 0000-0002-8721-6247\\
Michele Parrinello: 0000-0001-6550-3272\\

\noindent
{\bf Present Address}\\
$^\dagger$Department of Chemistry and Applied Biosciences, ETH
Zurich, c/o USI Campus, Via Giuseppe Buffi 13, CH-6900, Lugano, Ticino,
Switzerland\\
$^\ddagger$Facolt\`{a} di Informatica, Istituto di Scienze Computationali,
Universit\`{a} della Svizzera Italiana (USI), Via Giuseppe Buffi 13, CH-6900,
Lugano, Ticino, Switzerland.\\

\noindent
{\bf Notes:}\\ 
The authors declare no competing financial interest.\\

{\bf 7. ACKNOWLEDGEMENTS}\\

The authors would like to thank Luigi Bonati, Michele Invernizzi, Dr. Haiyang Niu, and Jayashrita Debnath for providing useful suggestions. We thank CSCS, Swiss National Supercomputing Centre for providing the computational resources. The research was supported by the European Union Grant No. ERC-2014-AdG-670227/VARMET. We also acknowledge the NCCR MARVEL, funded by the Swiss National Science Foundation.\\

\bibliography{ref}{}

\end{document}